\documentclass[pra,twocolumn,showpacs,preprintnumbers,superscriptaddress]
{revtex4}

\usepackage{times}
\usepackage{bm}
\usepackage{float}
\usepackage{graphicx}
\usepackage{amsbsy}
\usepackage{amsmath}
\usepackage{amsfonts}
\usepackage{amsthm}
\begin{document}

\theoremstyle{plain}
\newtheorem{theorem}{Theorem}
\newtheorem{lemma}[theorem]{Lemma}
\newtheorem{corollary}[theorem]{Corollary}
\newtheorem{proposition}[theorem]{Proposition}
\newtheorem{conjecture}[theorem]{Conjecture}

\theoremstyle{definition}
\newtheorem{definition}[theorem]{Definition}

\theoremstyle{remark}
\newtheorem*{remark}{Remark}
\newtheorem{example}{Example}
\title{Classification witness operator for the classification of different subclasses of three-qubit GHZ class}
\author{Anu Kumari, Satyabrata Adhikari}
\email{mkumari_phd2k18@dtu.ac.in, satyabrata@dtu.ac.in} \affiliation{Delhi Technological
University, Delhi-110042, Delhi, India}

\begin{abstract}
It is well known that three-qubit system has two kinds of inequivalent genuine entangled classes under stochastic local operation and classical communication (SLOCC). These classes are called as GHZ class and W class. GHZ class proved to be a very useful class for different quantum information processing tasks such as quantum teleportation, controlled quantum teleportation etc. In this work, we distribute pure three-qubit states from GHZ class into different subclasses denoted by $S_{1}$, $S_{2}$, $S_{3}$, $S_{4}$ and show that the three-qubit states either belong to   $S_{2}$ or $S_{3}$ or $S_{4}$ may be more efficient than the three-qubit state belong to $S_{1}$. Thus, it is necessary to discriminate the states belong to $S_{i}, i=2,3,4$ and the state belong to $S_{1}$. To achieve this task, we have constructed different witness operators that can classify the subclasses $S_{i}, i=2,3,4$ from $S_{1}$. We have shown that the constructed witness operator can be decomposed into Pauli matrices and hence can be realized experimentally.
\end{abstract}
\pacs{03.67.Hk, 03.67.-a} \maketitle

\section{Introduction}
Entanglement is a purely quantum mechanical phenomenon that plays a vital role in the advancement of quantum information theory. The two basic problems of quantum information theory are: (i) detection of n-qubit entangled states and (ii) classification of n-qubit entangled states. For $n=2$ i.e. for two-qubit quantum states, the only possibilities for the existence of quantum states are either as separable or entangled states. But as we increase the number of qubits, the complexity of the system will also increase. In these complex systems, the entangled states can be further classified as separable, biseparable, triseparable, genuine etc. If the entangled state is a genuine entangled state then it is entangled with respect to any partition.\\
Lot of research had already been done on the classification of entanglement. The problem on classification of entanglement started with the classification of three qubit pure states and it has been studied in the seminal work by Dur et.al. \cite{dur}. They have shown that three qubit pure states can be classified into six inequivalent classes under SLOCC: One separable state, three biseparable states and two genuinely entangled states. The two SLOCC inequivalent genuine entangled classes are GHZ class and W class. In the literature, it has been shown that there exist observables that can be used to distinguish the above mentioned six inequivalent classes of three-qubit pure states \cite{datta}. The experiment using NMR quantum information processor has been carried out to classify six inequivalent classes under SLOCC \cite{singh}. Acin et. al \cite{acin} have constructed witness operator to classify mixed three-qubit states. Sabin et. al. \cite{sabin} have studied the classification of pure as well as mixed three-qubit entanglement based on reduced two-qubit entanglement. Monogamy score can also be used to classify pure tripartite system \cite{bera}. The classification of different classes of four qubit pure states has been studied in \cite{verstraete,viehmann,zangi}. The number of different classes of n-qubit system increases when we increases the number of qubits. The discrimination of different classes of multi-qubit system has been studied in \cite{miyake,chen,li1,miyake1}.\\
In this work, we are focusing on the classification of the subclasses of GHZ class. To define different subclasses of GHZ class, let us consider the five parameter canonical form of three-qubit pure state $|\psi \rangle_{ABC}$ shared between three distant partners $A$, $B$ and $C$, which is given by \cite{acin1}
\begin{eqnarray}
|\psi \rangle_{ABC}&=&\lambda_0|000\rangle+\lambda_1e^{i\theta}|100\rangle+\lambda_2|101\rangle+\lambda_3|110\rangle\nonumber\\&+&\lambda_4|111\rangle
\label{canonical}
\end{eqnarray}
with $0\leq \lambda_{i}\leq 1 (i=0,1,2,3,4)$ and $0\leq \theta \leq \pi$.\\
The normalization condition of the state (\ref{canonical}) is given by
\begin{eqnarray}
\lambda_{0}^{2}+\lambda_{1}^{2}+\lambda_{2}^{2}+\lambda_{3}^{2}+\lambda_{4}^{2}=1.
\label{normalization}
\end{eqnarray}
The three-tangle $\tau_{\psi}$ for a pure three-qubit state $|\psi \rangle_{ABC}$ can be defined as \cite{coffman}
\begin{eqnarray}
\tau_{\psi}= C_{A(BC)}^{2}-C_{AB}^{2}-C_{AC}^{2}
\label{tangledef}
\end{eqnarray}
where $C_{AB}$, $C_{AC}$ represent the partial concurrences between the pairs $(A,B)$, $(A,C)$ respectively and $C_{A(BC)}$ denote the entanglement of qubit $A$ with the joint state of qubits $B$ and $C$. It can be interpreted as residual entanglement\cite{coffman}, which is not captured by two-qubit entanglement.\\
For a pure three-qubit state $|\psi \rangle_{ABC}$, The tangle $\tau_{\psi}$ can be calculated as\cite{datta}
\begin{eqnarray}
\tau_{\psi}= 4\lambda_{0}^{2}\lambda_{4}^{2}
\label{tanglecal}
\end{eqnarray}
The tangle $\tau_{\psi}\neq 0$ for GHZ class and $\tau_{\psi}= 0$ for W class of states. To define the subclasses of GHZ class, we assume that the state parameters $\lambda_{0}$ and $\lambda_{4}$ are not equal to zero. In this work, we will study the classification problem for the particular class of states in which the phase factor $\theta=0$. But similar calculations can be performed by taking $\theta \neq 0$ also.\\
We are now in a position to divide the three-qubit pure GHZ class of states (\ref{canonical}) into four subclasses as:
\begin{eqnarray}
\underline{\textbf{Subclass-I}:}\nonumber\\&&
S_{1}=\{|\psi_{S}\rangle\}, \textrm{where}\nonumber\\&&
|\psi_{S}\rangle = \lambda_0|000\rangle+\lambda_4|111\rangle
\label{class1}
\end{eqnarray}
\begin{eqnarray}
&&\underline{\textbf{Subclass-II}:}\nonumber\\&&
S_{2}=\{|\psi_{\lambda_{1}}\rangle,|\psi_{\lambda_{2}}\rangle,|\psi_{\lambda_{3}}\rangle\},\textrm{where}\nonumber\\&&
|\psi_{\lambda_{1}}\rangle=\lambda_0|000\rangle+\lambda_1|100\rangle+\lambda_4|111\rangle,\nonumber\\&&
|\psi_{\lambda_{2}}\rangle=\lambda_0|000\rangle+\lambda_2|101\rangle +\lambda_4|111\rangle,\nonumber\\&&
|\psi_{\lambda_{3}}\rangle=\lambda_0|000\rangle+\lambda_3|110\rangle +\lambda_4|111\rangle\}
\label{class2}
\end{eqnarray}
\begin{eqnarray}
&&\underline{\textbf{Subclass-III}:}\nonumber\\&&
S_{3}=\{|\psi_{\lambda_{1},\lambda_{2}}\rangle,|\psi_{\lambda_{1},\lambda_{3}}\rangle,|\psi_{\lambda_{2},\lambda_{3}}\rangle\},\textrm{where}\nonumber\\&&
|\psi_{\lambda_{1},\lambda_{2}}\rangle=\lambda_0|000\rangle+\lambda_1|100\rangle+\lambda_2|101\rangle+\lambda_4|111\rangle,\nonumber\\&& |\psi_{\lambda_{1},\lambda_{3}}\rangle=\lambda_0|000\rangle+\lambda_1|100\rangle+\lambda_3|110\rangle+\lambda_4|111\rangle,\nonumber\\&&
|\psi_{\lambda_{2},\lambda_{3}}\rangle=\lambda_0|000\rangle+\lambda_2|101\rangle+\lambda_3|110\rangle+\lambda_4|111\rangle
\label{class3}
\end{eqnarray}
\begin{eqnarray}
&&\underline{\textbf{Subclass-IV $(S_{4})$}:}\nonumber\\&&
S_{4}=\{|\psi_{\lambda_1,\lambda_2,\lambda_3}\rangle\}, \textrm{where}\nonumber\\&&
|\psi_{\lambda_1,\lambda_2,\lambda_3}\rangle= \lambda_0|000\rangle+\lambda_1|100\rangle+\lambda_2|101\rangle+\lambda_3|110\rangle+\nonumber\\&&\lambda_4|111\rangle
\label{class4}
\end{eqnarray}
Different subclasses of GHZ class of states are distributed in four different sets $S_{1}$, $S_{2}$, $S_{3}$, $S_{4}$. Classification of these subclasses  can be diagrammatically shown in Figure-I. In the Figure-I, the outermost circle represents GHZ states belonging to subclass-IV, the second outermost circle represent the GHZ states belonging to subclass-III, the third outermost circle represents the GHZ states belonging to subclass-II and the innermost circle represents the standard GHZ class of states belonging to subclass-I. We should note here that these subclasses are not inequivalent under SLOCC. To transform a state from one subclass to another, we need to perform local quantum operations that depend on the state which is to be transformed. So it is necessary to know the state or at least the subclass in which the state belongs. In this work, we would detect the subclass in which the state belongs.\\
\begin{figure}[h]
	\centering
	\includegraphics[scale=0.3]{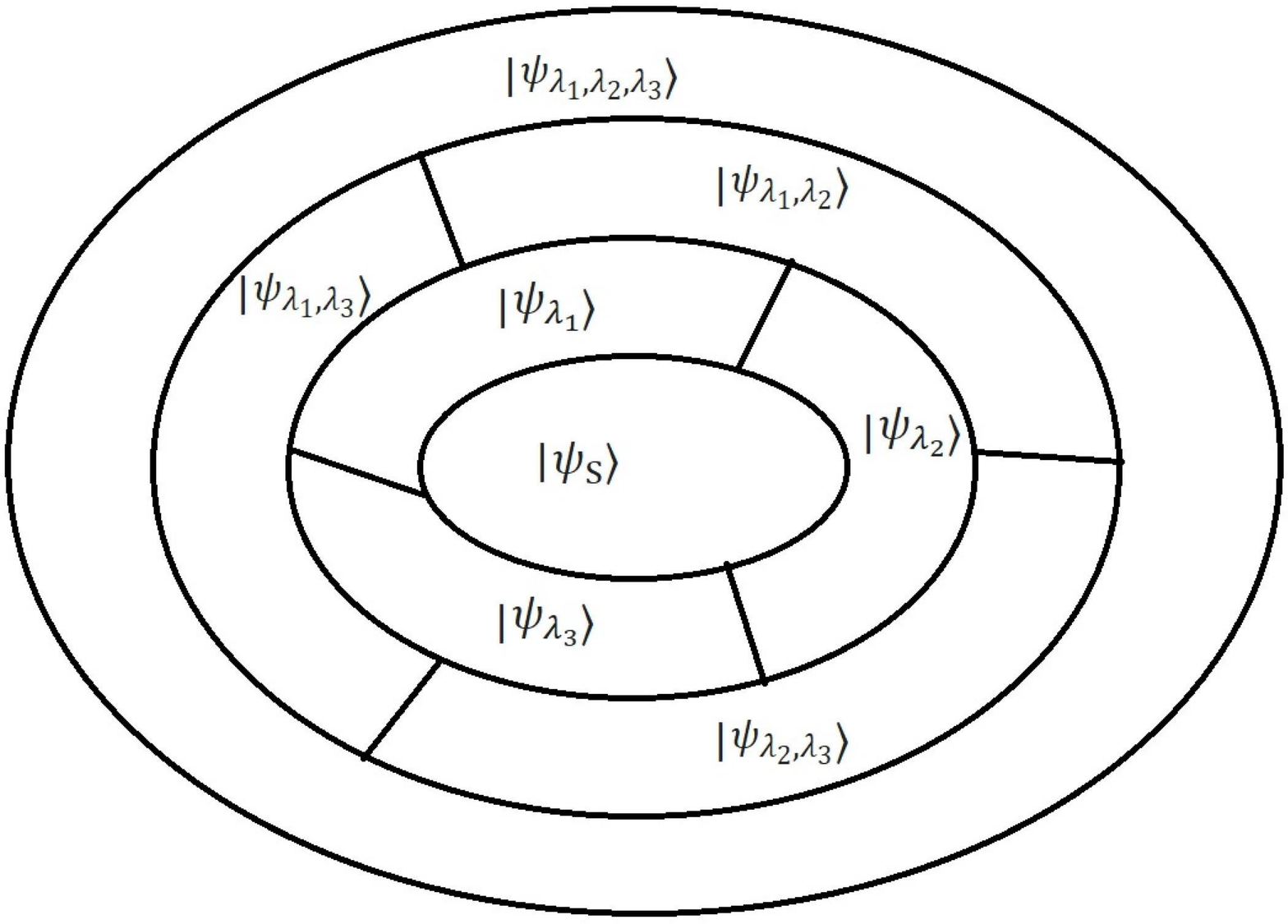}
	\caption{Classification of different subclasses of GHZ class of states described by the four sets $S_{1},S_{2}, S_{3}, S_{4}$}
\end{figure}
The motivation of the work is as follows: Firstly, let us consider the teleportation scheme introduced by Lee et.al. \cite{soojoonlee}. According to this teleportation scheme, a single-qubit measurement has been performed either on the qubit $A$ or qubit $B$ or qubit $C$ of the pure three-qubit state. After the measurement, the pure three-qubit state reduces to a two-qubit state at the output. Then the resulting two-qubit state  can be used as a resource state for quantum teleportation. The efficiency of the resource state is provided by the teleportation fidelity.\\
In particular, if the single-qubit measurement is performed on either qubit $A$ or qubit $B$ or qubit $C$ of the state $|\psi_{S}\rangle \in S_{1}$ then the corresponding maximal teleportation fidelities are given by \cite{soojoonlee}
\begin{eqnarray}
F_{A}^{(|\psi_{S}\rangle)}=F_{B}^{(|\psi_{S}\rangle)}=F_{C}^{(|\psi_{S}\rangle)}=\frac{2(1+\lambda_0\lambda_4)}{3}
\label{ghz1}
\end{eqnarray}
In a similar fashion, if the single-qubit measurement is performed on the state $|\psi_{\lambda_{1}^{'}}\rangle \in S_{2}$ then the corresponding maximal teleportation fidelities are given by \cite{soojoonlee}
\begin{eqnarray}
F_A^{(|\psi_{\lambda_{1}^{'}}\rangle )}&=&\frac{2(1+\lambda_4^{'}\sqrt{(\lambda_{0}^{'})^{2}+(\lambda_{1}^{'})^{2})}}{3}\nonumber\\
F_B^{(|\psi_{\lambda_{1}^{'}}\rangle )}&=&F_C^{(|\psi_{\lambda_{1}^{'}}\rangle )}=\frac{2(1+\lambda_{0}^{'}\lambda_{4}^{'})}{3}
\label{ghz2}
\end{eqnarray}
Again, if the single-qubit measurement is performed on the state $|\psi_{\lambda_{1}^{''}}\rangle \in S_{3}$ then the corresponding maximal teleportation fidelities are 
\begin{eqnarray}
F_A^{(|\psi_{\lambda_{1,2}^{''}}\rangle )}&=&\frac{2(1+\lambda_4^{''}\sqrt{(\lambda_{0}^{''})^{2}+(\lambda_{1}^{''})^{2})}}{3}\nonumber\\
F_B^{(|\psi_{\lambda_{1,2}^{''}}\rangle )}&=&\frac{2(1+\lambda_0^{''}\sqrt{(\lambda_{2}^{''})^{2}+(\lambda_{4}^{''})^{2})}}{3}\nonumber\\
F_C^{(|\psi_{\lambda_{1,2}^{''}}\rangle )}&=&\frac{2(1+\lambda_{0}^{''}\lambda_{4}^{''})}{3}
\label{ghz3}
\end{eqnarray}
and if the single-qubit measurement is performed on the state $|\psi_{\lambda_{1}^{'''}}\rangle \in S_{4}$ then the corresponding maximal teleportation fidelities are 
\begin{eqnarray}
F_A^{(|\psi_{\lambda_{1,2,3}^{'''}}\rangle )}&=&\frac{2(1+\sqrt{y})}{3}\nonumber\\
F_B^{(|\psi_{\lambda_{1,2,3}^{'''}}\rangle )}&=&\frac{2(1+\lambda_0^{'''}\sqrt{(\lambda_{2}^{'''})^{2}+(\lambda_{4}^{'''})^{2})}}{3}\nonumber\\
F_C^{(|\psi_{\lambda_{1,2,3}^{'''}}\rangle )}&=&\frac{2(1+\lambda_{0}^{'''}\sqrt{(\lambda_{3}^{'''})^{2}+(\lambda_{4}^{'''})^{2})}}{3}
\label{ghz4}
\end{eqnarray} 
where
\begin{eqnarray}
y=(\lambda_{0}^{'''})^{2}(\lambda_{4}^{'''})^{2}+(\lambda_{1}^{'''})^{2}(\lambda_{4}^{'''})^{2}+(\lambda_{2}^{'''})^{2}(\lambda_{3}^{'''})^{2}-4\lambda_{1}^{'''}\lambda_{2}^{'''}\lambda_{3}^{'''}\lambda_{4}^{'''}
\end{eqnarray}
It can be easily seen that there exist state parameters $\lambda_{0}^{'''},\lambda_{1}^{'''},\lambda_{2}^{'''}, \lambda_{3}^{'''}, \lambda_{4}^{'''}, \lambda_{0}^{''},\lambda_{1}^{''},\lambda_{2}^{''}, \lambda_{4}^{''}, \lambda_{0}^{'},\lambda_{1}^{'}, \lambda_{4}^{'}, \lambda_{0}$ and $\lambda_{4}$ such that the inequalities
\begin{eqnarray}
F_A^{(|\psi_{\lambda_{1,2}^{''}}\rangle )}&\geq& F_{A}^{(|\psi_{\lambda_{1}^{'}}\rangle)}, F_A^{(|\psi_{\lambda_{1,2,3}^{'''}}\rangle )}\geq F_{A}^{(|\psi_{\lambda_{1,2}^{''}}\rangle)}\nonumber\\ F_A^{(|\psi_{\lambda_{1,2,3}^{'''}}\rangle )}&\geq& F_{A}^{(|\psi_{\lambda_{1}^{'}}\rangle)},
F_A^{(|\psi_{\lambda_{1}^{'}}\rangle )}\geq F_{A}^{(|\psi_{S}\rangle)} \nonumber\\ F_A^{(|\psi_{\lambda_{1,2}^{''}}\rangle )}&\geq& F_{A}^{(|\psi_{S}\rangle)}, F_A^{(|\psi_{\lambda_{1,2,3}^{'''}}\rangle )}\geq F_{A}^{(|\psi_{S}\rangle)}
\label{compfid}
\end{eqnarray}
holds. In this way we can compare the teleportation fidelities of the GHZ states belonging to different subclasses.
We can conclude from (\ref{compfid}) that the pure three-qubit state $|\psi_{\lambda_{1}^{'}}\rangle \in S_{2}$ is more efficient than $|\psi_{S}\rangle \in S_{1}$ in the teleportation scheme \cite{soojoonlee}. In the same way, we can say that the states belonging to subclass $S_3$ are more efficient than the states belonging to $S_2$ or $S_1$. Also, it can be observed that the states belonging to any of the defined subclasses are GHZ states. Thus it is necessary to discriminate the pure three-qubit states belong to different subclasses of GHZ class.\\
Secondly,  we can compare the entanglement and the tangle in these subclasses.\\
(i) We can compare the entanglement between the reduced two qubit mixed states obtained after tracing out either subsystem A or subsystem B or subsystem C in the following way:\\
If we have GHZ state belonging to subclass $S_1$, then after tracing out one qubit, the concurrence of the resulting two qubit system will become zero, that is, $C_{AB}=C_{AC}=C_{BC}=0$. Thus, after tracing out one subsystem, the remaining two qubit state will become a separable state. Now, if we consider GHZ state belonging to subclass $S_2$, then we have exactly one of the concurrences either $C_{AB}$ or $C_{AC}$ or $C_{BC}$ of the mixed reduced system is non-zero. Thus, if we require any two qubit entangled state in some quantum information processing protocol, then we can obtain it by tracing out one qubit from three qubit GHZ state belong to subclass $S_2$. For example if we need any two qubit shared entangled state between Alice and Bob, then we can use three qubit GHZ state$(|\psi_{ABC}\rangle=\lambda_0|000\rangle+\lambda_3|110\rangle +\lambda_4|111\rangle)$, lying in subclass $S_2$.It is possible, since, the concurrence of the reduced state $\rho_{AB}=Tr_C(|\psi\rangle_{ABC}\langle \psi|)$ is not equal to zero. But this type of situation will not arise in the case of three qubit GHZ state belong to subclass $S_1$. Not only the subclass $S_2$, but we can use other subclasses such as $S_3$ and $S_4$ to get the entangled mixed two qubit state.\\
(ii) We can see changes in  tangle in these subclasses as follows:\\
 For a GHZ state belonging to $S_1$, we have only two parameters $\lambda_0$ and $\lambda_4$. But for the GHZ state belonging to $S_2$ have parameters $\lambda_0^{'}$, $\lambda_1^{'}$ and $\lambda_4^{'}$. Due to normalization condition, the values of parameters gets distributed.Thus, using normalization condition we get, $\lambda_0\lambda_4 > \lambda_0^{'}\lambda_4^{'}$. Since, tangle is defined as $\tau=4\lambda_0^{2}\lambda_4^{2}$, tangle of the three qubit GHZ state belonging to $S_1$ will be more then the tangle of the GHZ state belonging to $S_2$. Again, if we compare tangle of the three qubit GHZ state belonging to subclass $S_3$ will be more than the tangle of the GHZ state belonging to $S_2$.

In this way, we can conclude that
\begin{eqnarray}
\tau^{(2)} \geq \tau^{(3)} \geq \tau^{(4)} \geq \tau^{(5)}
\end{eqnarray}
where, $\tau^{(2)}$ is the tangle of the GHZ state belonging to sublclass $S_1$, $\tau^{(3)}$ is the tangle of the GHZ state belonging to subclass $S_2$, $\tau^{(4)}$ is the tangle of the GHZ state belonging to subclass $S_3$ and $\tau^{(5)}$ is the tangle of the GHZ state belonging to subclass $S_4$.
These are the few things that motivated us to classify different subclasses of GHZ states.\\
This paper is organized as follows: In Sec. II, we have revisited the correlation tensor for the canonical form of three-qubit pure state which will be needed in the later section. In Sec. III, we have constructed witness operator that can detect different subclasses of three -qubit pure GHZ class of states. In Sec. IV, we have verified our result with some examples. We conclude in Sec. V.
\section{Derivation of the inequality required for the construction of classification witness operator}
In this section, we will construct the Hermitian matrices from the component of the correlation tensor and then use its minimum and maximum eigenvalues to derive the required inequality for the construction of classification witness operator.\\
To start with, let us consider any arbitrary three qubit state described by the density operator $\rho$. The correlation coefficient of the state $\rho$ can be obtained as
\begin{eqnarray}
t_{ijk}=\text{Tr}(\rho(\sigma_i \otimes \sigma_j \otimes \sigma_k)), (i,j,k=x,y,z)
\end{eqnarray}
Then the correlation tensor $\vec{T}$ can be defined as $\vec{T}=(T_x,T_y,T_z)$, where
\begin{eqnarray}
T_x=
\begin{pmatrix}
t_{xxx} & t_{xyx} & t_{xzx}\\
t_{xxy} & t_{xyy} & t_{xzy}\\
t_{xxz} & t_{xyz} & t_{xzz}\\
\end{pmatrix}
\end{eqnarray}
and
\begin{eqnarray}
T_y=
\begin{pmatrix}
t_{yxx} & t_{yyx} & t_{yzx}\\
t_{yxy} & t_{yyy} & t_{yzy}\\
t_{yxz} & t_{yyz} & t_{yzz}\\
\end{pmatrix}
\end{eqnarray}
and
\begin{eqnarray}
T_z=
\begin{pmatrix}
t_{zxx} & t_{zyx} & t_{zzx}\\
t_{zxy} & t_{zyy} & t_{zzy}\\
t_{zxz} & t_{zyz} & t_{zzz}\\
\end{pmatrix}
\end{eqnarray}
\subsection{Correlation tensor for the canonical form of three-qubit pure state}
Let us consider the three-qubit pure state described by the density operator $\rho_{\psi}$
\begin{eqnarray}
\rho_{\psi}=|\psi \rangle_{ABC} \langle \psi|
\end{eqnarray}
where $|\psi \rangle_{ABC}$ is given by (\ref{canonical}).\\
The components $T_{x},T_{y}$ and $T_{z}$ of the correlation tensor $\vec{T}$ for the state $\rho_{\psi}$ is given by
\begin{eqnarray}
T_x=
\begin{pmatrix}
2\lambda_0\lambda_4 & 0 & 2\lambda_0\lambda_2\\
0 & -2\lambda_0\lambda_4 & 0\\
2\lambda_0\lambda_3 & 0 & 2\lambda_0\lambda_1 cos\theta\\
\end{pmatrix}
\label{tx}
\end{eqnarray}
\begin{eqnarray}
T_y=
\begin{pmatrix}
0 & -2\lambda_0\lambda_4 & 0\\
-2\lambda_0\lambda_4 & 0 & -2\lambda_0\lambda_2\\
0 & -2\lambda_0\lambda_3 & 2\lambda_0\lambda_1 sin\theta\\
\end{pmatrix}
\label{ty}
\end{eqnarray}
\begin{eqnarray}
T_z=
\begin{pmatrix}
t_{zxx} & t_{zyx} & t_{zzx}\\
t_{zxy} & t_{zyy} & t_{zzy}\\
t_{zxz} & t_{zyz} & t_{zzz}
\end{pmatrix}
\label{tz}
\end{eqnarray}
where $t_{zxx}=-2(\lambda_2\lambda_3+\lambda_1\lambda_4 cos\theta)$,$t_{zyx}=2\lambda_1\lambda_4 sin\theta$, $t_{zzx}=2(\lambda_3\lambda_4-\lambda_1\lambda_2 cos\theta)$,$t_{zxy}=2\lambda_1\lambda_4 sin\theta$, $t_{zyy}=2(\lambda_1\lambda_4 cos\theta-\lambda_2\lambda_3)$,$t_{zzy}=2\lambda_1\lambda_2 sin\theta$, $t_{zxz}=2(\lambda_2\lambda_4-\lambda_1\lambda_3 cos\theta)$, $t_{zyz}=2\lambda_1\lambda_3 sin\theta$, $t_{zzz}=\lambda_0^2-\lambda_1^2+\lambda_2^2+\lambda_3^2-\lambda_4^2$.\\
The Hermitian matrices can be constructed from $T_{x}$  and $T_{y}$ as
\begin{eqnarray}
T_x^{T}T_{x}=
\begin{pmatrix}
a_{x} & 0 & b_{x}\\
0 & c_{x} & 0\\
b_{x} & 0 & d_{x}\\
\end{pmatrix}
\label{hermitian2}
\end{eqnarray}
where $a_{x}=4\lambda_0^{2}(\lambda_4^{2}+\lambda_3^{2})$, $b_{x}=4\lambda_0^{2}(\lambda_2\lambda_4+\lambda_1\lambda_3 cos\theta)$,
$c_{x}=4\lambda_0^{2}\lambda_4^{2}$, $d_{x}=4\lambda_0^{2}(\lambda_2^{2}+\lambda_1^{2} cos^{2}\theta)$.\\
and
\begin{eqnarray}
T_y^{T}T_{y}=
\begin{pmatrix}
a_{y} & 0 & b_{y}\\
0 & c_{y} & d_{y}\\
b_{y} & d_{y} & e_{y}\\
\end{pmatrix}
\label{hermitian3}
\end{eqnarray}
where $a_{y}=4\lambda_0^{2}\lambda_4^{2}$, $b_{y}=4\lambda_0^{2}\lambda_2\lambda_4$, $c_{y}=4\lambda_0^{2}(\lambda_4^{2}+\lambda_3^{2})$,
$d_{y}=-4\lambda_0^{2}\lambda_1\lambda_3 sin\theta$, $e_{y}=4\lambda_0^{2}(\lambda_2^{2}+\lambda_1^{2} sin^{2}\theta)$.\\
The superscript $T$ refers to the simple matrix transposition operation.
\subsection{Inequality for the construction of classification witness operator}
Let us recall the canonical form of three-qubit state $|\psi\rangle_{ABC}$ given in (\ref{canonical}). The invariants with respect to the state $|\psi\rangle_{ABC}$ under local unitary transformations are given by \cite{adhikari}
\begin{eqnarray}
&&\lambda_{0}\lambda_{4}=\frac{\sqrt{\tau_{\psi}}}{2}\nonumber\\&&
\lambda_{0}\lambda_{2}=\frac{C_{AC}}{2}\nonumber\\&&
\lambda_{0}\lambda_{3}=\frac{C_{AB}}{2}\nonumber\\&&
|\lambda_{2}\lambda_{3}-e^{i\varphi}\lambda_{1}\lambda_{4}|=\frac{C_{BC}}{2}
\label{invariant}
\end{eqnarray}
Here $\tau_{\psi}$ denote the three-tangle of the state $|\psi\rangle_{ABC}$ whereas $C_{AB}$, $C_{AC}$ and $C_{BC}$ represent the partial concurrences between the pairs $(A,B)$, $(A,C)$ and $(B,C)$ respectively.\\
Furthermore, the invariants of three-qubit states under local unitary transformations has been studied in \cite{sudbery} and the invariants are given by
\begin{eqnarray}
&&I_1=\langle \psi|\psi\rangle\nonumber\\&&
I_2=tr(\rho_C^{2})=2(\lambda_1\lambda_2+\lambda_3\lambda_4)^2\nonumber\\&&
I_3=tr(\rho_B^{2})=2(\lambda_1\lambda_3+\lambda_2\lambda_4)^2\nonumber\\&&
I_4=tr(\rho_A^{2})=2\lambda_0^2\lambda_1^2\nonumber\\&&
I_5=\frac{1}{4}\tau_{\psi}^{2}=4\lambda_0^{4}\lambda_4^{4}
\end{eqnarray}
where $\rho_{A}=Tr_{BC}(|\psi\rangle_{ABC}\langle \psi|)$, $\rho_{B}=Tr_{AC}(|\psi\rangle_{ABC}\langle \psi|)$, $\rho_{C}=Tr_{AB}(|\psi\rangle_{ABC}\langle \psi|)$ denote reduced density matrices of a single qubit.\\
Further, recalling the Hermitian matrices $T_{x}^{T}T_{x}$ and $T_{y}^{T}T_{y}$ from (\ref{hermitian2}) and (\ref{hermitian3}), we calculate the traces of the Hermitian matrices as
\begin{eqnarray}
Tr({T_x}^{T}T_x)&=&8{\lambda_0}^2{\lambda_4}^2+4{\lambda_0}^2{\lambda_3}^2+4{\lambda_0}^2{\lambda_2}^2
\nonumber\\&+&4{\lambda_0}^2{\lambda_1}^2cos^{2}\theta
\label{Tr1}
\end{eqnarray}
and
\begin{eqnarray}
Tr({T_y}^{T}T_y)&=&8{\lambda_0}^2{\lambda_4}^2+4{\lambda_0}^2{\lambda_3}^2+4{\lambda_0}^2{\lambda_2}^2
\nonumber\\&+&4{\lambda_0}^2{\lambda_1}^2sin^{2}\theta
\label{Tr2}
\end{eqnarray}
Adding (\ref{Tr1}) and (\ref{Tr2}), we get
\begin{eqnarray}
Tr[({T_x}^{T}T_x)+({T_y}^{T}T_y)]&=&16{\lambda_0}^2{\lambda_4}^2+8{\lambda_0}^2{\lambda_3}^2\nonumber\\&+&8{\lambda_0}^2{\lambda_2}^2+4{\lambda_0}^2{\lambda_1}^2
\label{Tracecond1}
\end{eqnarray}
The expression for $Tr[({T_x}^{T}T_x)+({T_y}^{T}T_y)]$ can be re-expressed in terms of three-tangle and partial concurrences as
\begin{eqnarray}
Tr[({T_x}^{T}T_x)+({T_y}^{T}T_y)]&=&4\tau_{\psi}+2{C_{AB}}^2+2{C_{AC}}^2\nonumber\\&+&4{\lambda_0}^2{\lambda_1}^2
\label{Tracecond2}
\end{eqnarray}
In terms of expectation of the operators, the expression (\ref{Tracecond2}) can further be written as
\begin{eqnarray}
Tr[({T_x}^{T}T_x)+({T_y}^{T}T_y)]&=&(\langle O_1\rangle_{\psi_{ABC}})^2+\frac{1}{2}(\langle O_3\rangle_{\psi_{ABC}})^2\nonumber\\&+&\frac{1}{2}(\langle O_2 \rangle_{\psi_{ABC}})^2+4{\lambda_0}^2{\lambda_1}^2
\label{Tracecond3}
\end{eqnarray}
where
\begin{eqnarray}
O_1&=&2(\sigma_x \otimes \sigma_x \otimes \sigma_x)\nonumber\\
O_2&=&2(\sigma_x \otimes \sigma_z \otimes \sigma_x)\nonumber\\
O_3&=&2(\sigma_x \otimes \sigma_x \otimes \sigma_z)
\end{eqnarray}
The expection values of there operator may be written in terms of invariants as\cite{datta},
\begin{eqnarray}
\langle O_1\rangle =4\lambda_0\lambda_4=2\sqrt{\tau_{\psi}}\nonumber\\
\langle O_2\rangle =4\lambda_0\lambda_2=\frac{C_{AC}}{2}\nonumber\\
\langle O_3\rangle =4\lambda_0\lambda_3=\frac{C_{AB}}{2}
\end{eqnarray}
The upper bound (U) and the lower bound (L) of $Tr[({T_x}^{T}T_x)+({T_y}^{T}T_y)]$ is given by
\begin{eqnarray}
L \leq Tr[({T_x}^{T}T_x)+({T_y}^{T}T_y)]\leq U
\label{bound}
\end{eqnarray}
where $L=\mu_{max}({T_x}^{T}T_x)+\mu_{min}({T_y}^{T}T_y)$ and $U=(4\lambda_0\lambda_4+2\sqrt{2}\lambda_0\lambda_3+2\sqrt{2}\lambda_0\lambda_2+2\lambda_0\lambda_1)^2$. The lower bound $L$ can be obtained using Weyl's result \cite{horn}. $\mu_{max}({T_x}^{T}T_x)$ and $\mu_{min}({T_y}^{T}T_y)$ denote the maximum and minimum eigenvalue of $T_x^{T}T_x$ and $T_y^{T}T_y$ respectively.\\
Thus, equation (\ref{bound}) can be re-written as
\begin{eqnarray}
&&[\mu_{max}({T_x}^{T}T_x)+\mu_{min}({T_y}^{T}T_y)]^{\frac{1}{2}} \leq \nonumber\\&& 4\lambda_0\lambda_4+2\sqrt{2}\lambda_0\lambda_3+2\sqrt{2}\lambda_0\lambda_2+2\lambda_0\lambda_1
\label{inequality}
\end{eqnarray}
If $0\leq \mu_{max}({T_x}^{T}T_x)+\mu_{min}({T_y}^{T}T_y)\leq 1$ then the inequality (\ref{inequality}) reduces to
\begin{eqnarray}
&&\mu_{max}({T_x}^{T}T_x)+\mu_{min}({T_y}^{T}T_y) \leq \nonumber\\&& 4\lambda_0\lambda_4+2\sqrt{2}\lambda_0\lambda_3+2\sqrt{2}\lambda_0\lambda_2+2\lambda_0\lambda_1
\label{inequality1}
\end{eqnarray}

The derived inequality (\ref{inequality1}) will be useful in constructing the Hermitian operators for the classification of states lies within the subclasses of GHZ class.
\section{Construction of classification witness operator}
Let $|\chi\rangle$ and $|\omega\rangle$ be any states belong to subclass-I ($S_{1}$) and subclass-i ($S_{i}$) (i=II,III,IV) respectively. The Hermitian operator $H$ is said to be classification witness operator if
\begin{eqnarray}
&& (a) Tr(H|\chi\rangle\langle \chi|) \geq 0, \forall~~ |\chi\rangle \in S_{1} \nonumber\\&&
(b) Tr(H|\omega\rangle\langle \omega|) < 0, \textrm{for at least one}~ |\omega\rangle \in S_{i}, \nonumber\\&&
 (i=II,III,IV)
\label{cwo1}
\end{eqnarray}
If the above condition holds then the classification witness operator $H$ classifies the states between (i) subclass-I and subclass-II (ii) subclass-I and subclass-III (iii) subclass-I and subclass-IV.\\
In this section, we will discuss the procedure of constructing the different classification witness operators that can classify the states residing in (i) subclass-I and subclass-II (ii) subclass-I and subclass-III (iii) subclass-I and subclass-IV.

\subsection{Classification witness operator for the classification of states contained in subclass-I and subclass-II}
We are now in a position to construct the classification witness operator that can classify the states resides in subclass-I and subclass-II.
\subsubsection{Classification of states confined in subclass-II with state parameters $\lambda_0$, $\lambda_1$ and $\lambda_4$ and subclass-I}
The GHZ class of state within subclass-II with state parameters $\lambda_0$, $\lambda_1$ and $\lambda_4$ is given by
\begin{eqnarray}
|\psi_{\lambda_1}\rangle=\lambda_0|000\rangle+\lambda_1|100\rangle+\lambda_4|111\rangle
\label{psi1}
\end{eqnarray}
with the normalization condition $\lambda_0^{2}+\lambda_1^{2}+\lambda_4^{2}=1$.\\
In particular, for $\lambda_1=0$, the state $|\psi_{\lambda_1}\rangle$ reduces to $|\psi_{\lambda_1=0}\rangle \in S_{1}$
where
\begin{eqnarray}
|\psi_{\lambda_1=0}\rangle=\lambda_0|000\rangle+\lambda_4|111\rangle, \lambda_0^{2}+\lambda_4^{2}=1
\label{psi0}
\end{eqnarray}
The Hermitian matrices $T_{x}^{T}T_{x}$ and $T_{y}^{T}T_{y}$ for the state $\rho_{\lambda_1}=|\psi_{\lambda_1 }\rangle \langle \psi_{\lambda_1}|$ is given by
\begin{eqnarray}
T_{x}^{T}T_{x}=
\begin{pmatrix}
4\lambda_0^{2}\lambda_4^{2} & 0 & 0\\
0 & 4\lambda_0^{2}\lambda_4^{2} & 0\\
0 & 0 & 4\lambda_0^{2}\lambda_1^{2}\\
\end{pmatrix}
\end{eqnarray}
\begin{eqnarray}
T_{y}^{T}T_{y}=
\begin{pmatrix}
4\lambda_0^{2}\lambda_4^{2} & 0 & 0\\
0 & 4\lambda_0^{2}\lambda_4^{2} & 0\\
0 & 0 & 0\\
\end{pmatrix}
\end{eqnarray}
The expression for $Tr[({T_x}^{T}T_x)+({T_y}^{T}T_y)]$ is given by
\begin{eqnarray}
Tr[({T_x}^{T}T_x)+({T_y}^{T}T_y)]&=&16{\lambda_0}^2{\lambda_4}^2+4{\lambda_0}^2{\lambda_1}^2
\label{Tracecond21}
\end{eqnarray}
The maximum eigenvalue of $T_{x}^{T}T_{x}$ is given by
\begin{eqnarray}
\mu_{max}({T_x}^{T}T_x)=max\{4\lambda_0^{2}\lambda_1^{2},4\lambda_0^{2}\lambda_4^{2}\}
\label{maxeigenval1}
\end{eqnarray}
The minimum eigenvalue of $T_{y}^{T}T_{y}$ is given by
\begin{eqnarray}
\mu_{min}({T_y}^{T}T_y)=0
\label{mineigenval1}
\end{eqnarray}
It can be easily observed that in this case $0\leq \mu_{max}({T_x}^{T}T_x)+\mu_{min}({T_y}^{T}T_y)\leq 1$ holds.\\
Since $\mu_{max}({T_x}^{T}T_x)$ depends on the value of the two parameters $\lambda_1$ and $\lambda_4$ so we will investigate two cases independently.\\
\underline{\textbf{Case-I: $\lambda_{4}>\lambda_{1}$}}\\
If $\lambda_{4}>\lambda_{1}$ then $\mu_{max}({T_x}^{T}T_x)=4\lambda_0^{2}\lambda_4^{2}$.\\
The inequality (\ref{inequality1}) then can be re-expressed in terms of the expectation value of the operators $O_{1}$ as
\begin{eqnarray}
-2\lambda_0\lambda_1 \leq \langle O_1\rangle_{\psi_{\lambda_1}}-\frac{\langle \langle O_1\rangle_{\psi_{\lambda_1}}\rangle^{2}}{4}
\label{inequality11}
\end{eqnarray}
If $\lambda_1=0$ then the R.H.S of the inequality (\ref{inequality11}) is always positive. Further, it can be observed that since $0\leq \langle O_1\rangle_{\psi_{\lambda_1}}\leq 1$ so the R.H.S of the inequality (\ref{inequality11}) still positive even for $\lambda_1 \neq 0$. Thus the R.H.S of the inequality is positive for every state belong to $S_{2}$. Hence, for $\lambda_{4}>\lambda_{1}$, it is not possible to make a distinction between the class of states $|\psi_{\lambda_{1}}\rangle \in S_{2} $ and $|\psi_{\lambda_{1}=0}\rangle \in S_{1}$ using the inequality (\ref{inequality11}).\\
\underline{\textbf{Case-II: $\lambda_{4}<\lambda_{1}$}}\\
If $\lambda_{4}<\lambda_{1}$ then $\mu_{max}({T_x}^{T}T_x)=4\lambda_0^{2}\lambda_1^{2}$.\\
The inequality (\ref{inequality1}) can be re-written as
\begin{eqnarray}
-2\lambda_0\lambda_1 \leq \langle O_1\rangle_{\psi_{\lambda_1}}-4\lambda_0^{2}\lambda_1^{2}
\label{inequality12}
\end{eqnarray}
We can now define an Hermitian operator $H_{1}$ as
\begin{eqnarray}
H_{1}=O_1-\frac{1}{4}\langle O_4\rangle_{\psi_{\lambda_1}}^{2}I
\label{h1}
\end{eqnarray}
where,
\begin{eqnarray}
O_4=2(\sigma_{x} \otimes I \otimes I)
\end{eqnarray}
The expectation value of the operator $O_4$, in terms of invariants may be written as,
\begin{eqnarray}
\langle O_4\rangle=4\lambda_0\lambda_1=4\sqrt{\frac{I_4}{\sqrt{2}}}
\end{eqnarray}
Therefore, the inequality (\ref{inequality12}) can be re-formulated as
\begin{eqnarray}
-2\lambda_0\lambda_1 \leq \langle H_{1}\rangle_{\psi_{\lambda_1}}
\label{inequality13}
\end{eqnarray}
If $\lambda_1=0$ then $\langle H_{1}\rangle_{\psi_{\lambda_1}}\geq 0$ for all states $|\psi_{\lambda_1=0}\rangle \in S_{1}$.\\
For $\lambda_1\neq0$, we can calculate $\langle H_{1}\rangle_{\psi_{\lambda_1}}
\label{inequality13}=Tr(H_{1}\rho_{\lambda_1})$ which is given by
\begin{eqnarray}
Tr(H_{1}\rho_{\lambda_1})&=&4\lambda_0(\lambda_4-\lambda_0{\lambda_1}^2)
\label{tr12}
\end{eqnarray}
It can be easily shown that there exist state parameters $\lambda_0,\lambda_1,\lambda_4$ for which $\lambda_4-\lambda_0{\lambda_1}^2<0$ and thus $Tr(H_{1}\rho_{\lambda_1})<0$. For instance, if we take $\lambda_0=0.4$, $\lambda_1=0.911043$ and $\lambda_4=0.1$, Then $Tr(H_1\rho_{\lambda_1})=-0.3712$, which is negative.\\
Thus the Hermitian operator $H_{1}$ discriminate the class $|\psi_{\lambda_{1}}\rangle \in S_{2}$ from $|\psi_{\lambda_1=0}\rangle \in S_{1}$.
\subsubsection{Classification of states confined in subclass-II with state parameters $\lambda_0$, $\lambda_i (i=2,3)$ and $\lambda_4$ and subclass-I}
The GHZ class of state within subclass-II with state parameters ($\lambda_0$, $\lambda_2$, $\lambda_4$) and ($\lambda_0$, $\lambda_3$, $\lambda_4$) are given by
\begin{eqnarray}
|\psi_{\lambda_2}\rangle=\lambda_0|000\rangle+\lambda_2|101\rangle+\lambda_4|111\rangle
\label{psi2}
\end{eqnarray}
with $\lambda_0^{2}+\lambda_2^{2}+\lambda_4^{2}=1$ and
\begin{eqnarray}
|\psi_{\lambda_3}\rangle=\lambda_0|000\rangle+\lambda_3|110\rangle+\lambda_4|111\rangle
\label{psi3}
\end{eqnarray}
with $\lambda_0^{2}+\lambda_3^{2}+\lambda_4^{2}=1$.\\
The Hermitian matrices $T_{x}^{T}T_{x}$ and $T_{y}^{T}T_{y}$ for the state $\rho_{\lambda_2}=|\psi_{\lambda_2 }\rangle \langle \psi_{\lambda_2}|$ and the state $\rho_{\lambda_3}=|\psi_{\lambda_3 }\rangle \langle \psi_{\lambda_3}|$ are given in appendix-1.

For the state either described by the density operator $\rho_{\lambda_2}=|\psi_{\lambda_2}\rangle \langle \psi_{\lambda_2}|$ or $\rho_{\lambda_3}=|\psi_{\lambda_3 }\rangle \langle \psi_{\lambda_3}|$, the expression of $Tr[({T_x}^{T}T_x)+({T_y}^{T}T_y)]$ is given by
\begin{eqnarray}
Tr[({T_x}^{T}T_x)+({T_y}^{T}T_y)]&=&16\lambda_0^{2}\lambda_4^{2}+8\lambda_0^{2}\lambda_i^{2},\\&& (i=2,3)\nonumber
\label{Tracecond21}
\end{eqnarray}

The inequality (\ref{inequality1}) then can be re-expressed in terms of the expectation value of the operators $O_{1}$ and $O_{4}$ as
\begin{eqnarray}
-2\sqrt{2}\lambda_0\lambda_i &\leq& \langle O_1\rangle_{\psi_{\lambda_i}}-\frac{1}{4}\langle O_1\rangle_{\psi_{\lambda_i}}^{2}-\frac{1}{4}\langle O_i\rangle_{\psi_{\lambda_i}}^{2},\\&&(i=2,3)\nonumber
\label{inequality21}
\end{eqnarray}
We can now define classification witness operators $H_{i}, (i=2,3)$ as
\begin{eqnarray}
H_{i}=O_1-\frac{1}{4}[\langle O_i\rangle_{\psi_{\lambda_i}}^{2}+\langle O_1\rangle_{\psi_{\lambda_i}}^{2}]I
\label{h2}
\end{eqnarray}
Therefore, the inequality (\ref{inequality21}) can be re-formulated as
\begin{eqnarray}
-2\sqrt{2}\lambda_0\lambda_i \leq \langle H_{i}\rangle_{\psi_{\lambda_i}}, i=2,3
\label{inequality22}
\end{eqnarray}
If $\lambda_i=0, (i=2,3) $ then $\langle H_{i}\rangle_{\psi_{\lambda_i}}\geq 0$ for all states $|\psi_{\lambda_i=0}\rangle (i=2,3) \in S_{1}$.\\
For $\lambda_i\neq0, (i=2,3)$, we can calculate $Tr(H_{i}\rho_{\lambda_i}) (i=2,3)$ which is given by
\begin{eqnarray}
Tr(H_{i}\rho_{\lambda_i})&=&4\lambda_0\lambda_4(1-\lambda_0\lambda_4)-4{\lambda_0}^2{\lambda_i}^2,i=2,3
\label{tr12}
\end{eqnarray}
It can be easily shown that there exist state parameters $\lambda_0,\lambda_i (i=2,3),\lambda_4$ for which $Tr(H_{i}\rho_{\lambda_i})<0$. For instance, if we  take $\lambda_0=0.4$,$\lambda_i$=0.894427$ (i=2,3)$ and $\lambda_4=0.2$, we get $Tr[H_i\rho_{\lambda_i}]=-0.2176$.
Therefore, the classifcation witness operator $H_{i} (i=2,3)$ classify the class of states $\psi_{\lambda_i} (i=2,3) \in S_{2}$ given in (\ref{psi2}) from the class $|\psi_{\lambda_i=0}\rangle (i=2,3) \in S_{1}$.\\
\subsection{Classification witness operator for the classification of states contained in subclass-I and subclass-III}
In this subsection, we will construct classification witness operator to discriminate subclass-I from subclasses of GHZ class spanned by four basis states.

\subsubsection{Classification of states confined in subclass-III with state parameters $\lambda_0$, $\lambda_2$, $\lambda_3$ and $\lambda_4$ and subclass-I}
The GHZ class of state within subclass-III with state parameters $\lambda_0$, $\lambda_2$, $\lambda_3$ and $\lambda_4$ is given by
\begin{eqnarray}
|\psi_{\lambda_{2},\lambda_{3}}\rangle=\lambda_0|000\rangle+\lambda_2|101\rangle+\lambda_3|110\rangle+\lambda_4|111\rangle
\label{psi23}
\end{eqnarray}
with $\lambda_0^{2}+\lambda_2^{2}+\lambda_3^{2}+\lambda_4^{2}=1$.\\
The Hermitian matrices $T_{x}^{T}T_{x}$ and $T_{y}^{T}T_{y}$ for the state $\rho_{\lambda_{2},\lambda_{3}}=|\psi_{\lambda_{2},\lambda_{3}}\rangle \langle \psi_{\lambda_{2},\lambda_{3}}|$ is given in appendix-2.

The expression for $Tr[({T_x}^{T}T_x)+({T_y}^{T}T_y)]$ is given by
\begin{eqnarray}
Tr[({T_x}^{T}T_x)+({T_y}^{T}T_y)]&=&16\lambda_0^{2}\lambda_4^{2}+8\lambda_0^{2}\lambda_2^{2}\nonumber\\&&+8\lambda_0^{2}\lambda_{3}^{2}
\label{Tracecond121}
\end{eqnarray}
\textbf{Case-I:} If $\mu_{max}({T_x}^{T}T_x)=u=4\lambda_0^{2}\lambda_4^{2}$ and $\mu_{min}({T_y}^{T}T_y)=0$.
The inequality (\ref{inequality1}) then can be re-expressed in terms of the expectation value of the operators $O_{1}$ as
\begin{eqnarray}
-2\sqrt{2}\lambda_0(\lambda_2+\lambda_3) &\leq& \langle O_1\rangle_{\psi_{\lambda_2,\lambda_3}}-\frac{\langle\langle O_1\rangle_{\psi_{\lambda_2,\lambda_3}}\rangle^{2}}{4}
\label{inequality231}
\end{eqnarray}
If $-2\sqrt{2}\lambda_0(\lambda_2+\lambda_3)=0$, then RHS of inquality (\ref{inequality231}) is always positive for every state $|\psi \rangle$. Thus, in this case, it is not possible to discriminate between the class of states $|\psi_{\lambda_{2}=0,\lambda_{3}=0}\rangle \in S_{1}$ and the class of  states $|\psi_{\lambda_{2},\lambda_{3}}\rangle \in S_{3}$.\\

\textbf{Case-II:}
If $\mu_{max}({T_x}^{T}T_x)= v_{1}$ and $\mu_{min}({T_y}^{T}T_y)=0$
then the inequality (\ref{inequality1}) reduces to
\begin{eqnarray}
-2\sqrt{2}\lambda_0(\lambda_2+\lambda_3) &\leq& \langle O_1\rangle_{\psi_{_{\lambda_2,\lambda_3}}}-P_1
\label{inequality232}
\end{eqnarray}
where,
\begin{eqnarray}
P_1&=&2\lambda_0^{2}(1-\lambda_0^2+\sqrt{-4\lambda_2^2\lambda_3^{2}+(1-\lambda_0^2)^2})\nonumber\\&=&2\langle O_5\rangle_{\psi_{\lambda_2,\lambda_3}}(1-\langle O_5\rangle_{\psi_{\lambda_2,\lambda_3}}\nonumber\\&+&\sqrt{-\frac{1}{4}\langle O_6\rangle_{\psi_{\lambda_2,\lambda_3}}^{2}+(1-\langle O_5\rangle_{\psi_{\lambda_2,\lambda_3}})^{2})}
\end{eqnarray}
where,
\begin{eqnarray}
O_5&=&\frac{1}{8}(I+\sigma_{z} \otimes I+\sigma_{z} \otimes I+\sigma_{z})\nonumber\\
O_6&=& 2(I \otimes \sigma_{y} \otimes \sigma _{y})
\end{eqnarray}
The expectation values of the operators $O_5$ and $O_6$, in terms of invariants may be written as,
\begin{eqnarray}
\langle O_5\rangle&=&\lambda_0^{2}=\frac{C_{AC}C_{AB}}{2C_{BC}}-\frac{1}{C_{BC}}\sqrt{\frac{2I_{4}I_{5}}{\tau_{\psi}}}\nonumber\\
\langle O_6\rangle&=&4(\lambda_2\lambda_3-\lambda_1\lambda_4)=2C_{BC}
\end{eqnarray}
We can now define an Hermitian operator $H_{5}$ as
\begin{eqnarray}
H_{4}&=&O_1-P_1I
\label{h4}
\end{eqnarray}
Therefore, the inequality (\ref{inequality232}) can be re-formulated as
\begin{eqnarray}
-2\sqrt{2}\lambda_0(\lambda_2+\lambda_3) \leq \langle H_{4}\rangle_{\psi_{\lambda_2,\lambda_3}}
\label{inequality233}
\end{eqnarray}
If $\lambda_2+\lambda_3=0$ then $\langle H_{4}\rangle_{\psi_{\lambda_2,\lambda_3}} \geq 0$ for all states $|\psi_{\lambda_{2}=0,\lambda_{3}=0}\rangle$.\\
For $-2\sqrt{2}\lambda_0(\lambda_2+\lambda_3)\neq0$, we can calculate $Tr(H_{4}\rho_{\lambda_2,\lambda_{3}})$, which is given by
\begin{eqnarray}
Tr(H_{4}\rho_{\lambda_{2},\lambda_{3}})&=&4\lambda_0\lambda_4-2{\lambda_0}^2(1-{\lambda_0}^2+\sqrt{T_1})
\label{tr23}
\end{eqnarray}
where,
\begin{eqnarray}
T_1={\lambda_2}^4-2{\lambda_2}^2{\lambda_3}^2+2{\lambda_2}^2{\lambda_4}^2+({\lambda_3}^2+{\lambda_4}^2)^2
\end{eqnarray}
It can be easily shown that there exist state parameters $(\lambda_0,\lambda_2, \lambda_3, \lambda_4)$ for which $Tr(H_{4}\rho_{\lambda_{2},\lambda_{3}})<0$. For instance, if we take $\lambda_0=0.35$, $\lambda_2=0.3$, $\lambda_3=0.864581$ and $\lambda_4=0.2$, we get $Tr[H_4\rho_{\lambda_{2},\lambda_{3}}]=-0.108386$. Therefore, the classification operator $H_{4}$ classify the class of states $\rho_{\lambda_{2},\lambda_{3}}\in S_{3}$ and the class of states $\rho_{\lambda_{2}=0,\lambda_{3}=0}\in S_{1}$.

\subsubsection{Classification of states confined in subclass-III with state parameters $\lambda_0$, $\lambda_1$, $\lambda_i (i=2,3)$ and $\lambda_4$ and subclass-I}
The GHZ class of state within subclass-III with state parameters ($\lambda_0$, $\lambda_1$, $\lambda_2$, $\lambda_4$) and ($\lambda_0$, $\lambda_1$, $\lambda_3$, $\lambda_4$) are given by
\begin{eqnarray}
|\psi_{\lambda_{1},\lambda_{2}}\rangle=\lambda_0|000\rangle+\lambda_1|100\rangle+\lambda_2|101\rangle+\lambda_4|111\rangle
\label{psi102}
\end{eqnarray}
with $\lambda_0^{2}+\lambda_1^{2}+\lambda_2^{2}+\lambda_4^{2}=1$.
\begin{eqnarray}
|\psi_{\lambda_{1},\lambda_{3}}\rangle=\lambda_0|000\rangle+\lambda_1|100\rangle+\lambda_3|110\rangle+\lambda_4|111\rangle
\label{psi13}
\end{eqnarray}
with $\lambda_0^{2}+\lambda_1^{2}+\lambda_3^{2}+\lambda_4^{2}=1$.\\
The Hermitian matrices $T_{x}^{T}T_{x}$ and $T_{y}^{T}T_{y}$ for the state $\rho_{\lambda_{1},\lambda_{2}}=|\psi_{\lambda_{1},\lambda_{2}}\rangle \langle \psi_{\lambda_{1},\lambda_{2}}|$ and the state $\rho_{\lambda_{1,3}}=|\psi_{\lambda_{1,3}}\rangle \langle \psi_{\lambda_{1,3}}|$ are given in appendix-3.\\

The expression for $Tr[({T_x}^{T}T_x)+({T_y}^{T}T_y)]$ is given by
\begin{eqnarray}
Tr[({T_x}^{T}T_x)+({T_y}^{T}T_y)]&=&16\lambda_0^{2}\lambda_4^{2}+8\lambda_0^{2}\lambda_i^{2}+\nonumber\\&&
4\lambda_0^{2}\lambda_1^{2},~~(i=2,3)
\label{Tracecond131}
\end{eqnarray}

\textbf{Case-I:} If $\mu_{max}({T_x}^{T}T_x)=4\lambda_0^{2}\lambda_4^{2}$ and $\mu_{min}({T_y}^{T}T_y)=0$.
The inequality (\ref{inequality1}) then can be re-expressed in terms of the expectation value of the operators $O_{1}$ as
\begin{eqnarray}
-2\lambda_0(\sqrt{2}\lambda_i+\lambda_1) &\leq& \langle O_1\rangle_{\psi_{\lambda_1,\lambda_i}}-\frac{(\langle O_1\rangle_{\psi_{\lambda_1,\lambda_i}})^{2}}{4},~i=2,3
\label{inequality121}
\end{eqnarray}
If $\sqrt{2}\lambda_i+\lambda_1=0 (i=2,3)$, then RHS of inquality (\ref{inequality121}) is always positive. Thus, the R.H.S of the inequality is positive for every state $|\psi\rangle$.  But since $0\leq \langle O_1\rangle_{\psi_{\lambda_1,\lambda_i}}\leq 1$ so the R.H.S of the inequality (\ref{inequality121}) still positive even for $\sqrt{2}\lambda_i+\lambda_1 \neq 0, (i=2,3)$. Thus it is not possible to differentiate between the class of states $|\psi_{\lambda_{1}=0,\lambda_{i}=0}\rangle \in S_{1} (i=2,3)$ and $|\psi_{\lambda_{1},\lambda_{i}}\rangle \in S_{3} (i=2,3)$, using the inequality (\ref{inequality121}) for this case.\\
\textbf{Case-II:}
If $\mu_{max}({T_x}^{T}T_x)=2\lambda_0^{2}(\lambda_1^{2}+\lambda_{i}^{2}+\lambda_4^{2}+\sqrt{(\lambda_1^{2}+\lambda_{i}^{2}+\lambda_4^{2})^2-4\lambda_1^2\lambda_4^{2}})$,~i=2,3 and $\mu_{min}({T_y}^{T}T_y)=0$
Then the inequality (\ref{inequality1}) can be re-written as
\begin{eqnarray}
-2\lambda_0(\sqrt{2}\lambda_{i}+\lambda_1) &\leq& \langle O_1\rangle_{\psi_{\lambda_1,\lambda_i}}-P_{i},i=2,3
\label{inequality122}
\end{eqnarray}
where, for i=2,3
\begin{eqnarray}
P_i&=&2\lambda_0^{2}(\lambda_1^{2}+\lambda_{i}^{2}+\lambda_4^{2}\nonumber\\&+&\sqrt{(\lambda_1^{2}+\lambda_{i}^{2}+\lambda_4^{2})^2-4\lambda_1^2\lambda_4^{2}})
\end{eqnarray}
$P_{i}(i=2,3)$ can also be re-expressed in terms of the expectation values of the operators $O_1$, $O_4$ and $O_{i}(i=2,3)$ as
\begin{eqnarray}
P_i&=&q+\sqrt{q^2-\frac{1}{16}\langle O_1\rangle_{\psi_{\lambda_1,\lambda_i}}^2\langle O_4\rangle_{\psi_{\lambda_1,\lambda_i}}^2}
\end{eqnarray}
where $q=\frac{1}{8}[\langle O_1\rangle_{\psi_{\lambda_1,\lambda_i}}^{2}+\langle O_i\rangle_{\psi_{\lambda_1,\lambda_i}}^{2}+\langle O_4\rangle_{\psi_{\lambda_1,\lambda_i}}^{2}]$
We can now define an Hermitian operator $H_{k}, k=5,6$ as
\begin{eqnarray}
H_{k}&=&O_1-P_iI, ~k=5,6, i=2,3, i=k
\label{h5}
\end{eqnarray}
Therefore, the inequality (\ref{inequality122}) can be re-formulated as
\begin{eqnarray}
-2\lambda_0
(\sqrt{2}\lambda_i+\lambda_1) \leq \langle H_{k}\rangle_{\psi_{\lambda_1,\lambda_i}},~~i=2,3, k=5,6
\label{inequality123}
\end{eqnarray}
If $\sqrt{2}\lambda_{i}+\lambda_1=0,i=2,3$ then $\langle H_{k}\rangle_{\psi_{\lambda_1,\lambda_i}} \geq 0, k=5,6$ for all states $|\psi_{\lambda_{1}=0,\lambda_{i}=0}\rangle \in S_{1} (i=2,3)$.\\
For $\sqrt{2}\lambda_{i}+\lambda_1 \neq0, i=2,3$, we can calculate $Tr(H_{k}\rho_{\lambda_{1},\lambda_{i}}),(i=2,3, k=5,6)$, which is given by
\begin{eqnarray}
Tr(H_{k}\rho_{\lambda_{1},\lambda_{i}})&=&4\lambda_0\lambda_4-2{\lambda_0}^2(\lambda_1^2+\lambda_{i}^2+\lambda_4^{2}+\sqrt{T_{i}}),\\&&\nonumber i=2,3, k=5,6
\label{tr12}
\end{eqnarray}
where
\begin{eqnarray}
T_{i}={\lambda_1}^4+2{\lambda_1}^2({\lambda_{i}}^2-{\lambda_4}^2)+({\lambda_{i}}^2+{\lambda_4}^2)^2,i=2,3
\end{eqnarray}
It can be easily shown that there exist state parameters $(\lambda_0,\lambda_1, \lambda_{i}, \lambda_4), (i=2,3)$ for which $Tr(H_{k}\rho_{\lambda_1,\lambda_{i}})<0$ for k=5,6. For instance, if we take $\lambda_0=0.5$, $\lambda_1=0.83666$, $\lambda_i=0.2,(i=2,3)$ and $\lambda_4=0.1$, we get $Tr[H_k\rho_{\lambda_{1},\lambda_{i}}]=-0.540548, (k=5,6)$. Thus, the Hermitian operator $H_{k}$, k=\{5,6\} serves as a classification witness operator and classify the class of states described by the density operator $\rho_{\lambda_{1},\lambda_{i}},(i=2,3) \in S_{3}$ and the class of states $\rho_{\lambda_{1}=0,\lambda_{i}=0},(i=2,3) \in S_{1}$.


\subsection{Classification of states confined in subclass-IV with state parameters $(\lambda_0$, $\lambda_1$,$\lambda_2$ $\lambda_3$, $\lambda_4)$ and subclass-I}
The GHZ class of state within subclass-IV with state parameters ($\lambda_0$, $\lambda_1$, $\lambda_2$, $\lambda_3$, $\lambda_4$) is given by
\begin{eqnarray}
|\psi_{\lambda_{1},\lambda_{2},\lambda_{3}}\rangle&=&\lambda_0|000\rangle+\lambda_1|100\rangle+\lambda_2|101\rangle+\lambda_3|110\rangle \nonumber\\&+& \lambda_4|111\rangle
\label{psi123}
\end{eqnarray}
with $\lambda_0^{2}+\lambda_1^{2}+\lambda_2^{2}+\lambda_3^{2}+\lambda_4^{2}=1$.\\
The Hermitian matrices $T_{x}^{T}T_{x}$ and $T_{y}^{T}T_{y}$ for the state $\rho_{\lambda_{1},\lambda_{2},\lambda_{3}}=|\psi_{\lambda_{1},\lambda_{2},\lambda_{3}}\rangle \langle \psi_{\lambda_{1},\lambda_{2},\lambda_{3}}|$ is given in appendix-4.

The expression for $Tr[({T_x}^{T}T_x)+({T_y}^{T}T_y)]$ is given by
\begin{eqnarray}
Tr[({T_x}^{T}T_x)+({T_y}^{T}T_y)]&=&16\lambda_0^{2}\lambda_4^{2}+8\lambda_0^{2}\lambda_3^{2}+8\lambda_0^{2}\lambda_2^{2}\nonumber\\&&
+4\lambda_0^{2}\lambda_1^{2}
\label{Tracecond1231}
\end{eqnarray}
\textbf{Case-I:} If $\mu_{max}({T_x}^{T}T_x)=u=4\lambda_0^{2}\lambda_4^{2}$ and $\mu_{min}({T_y}^{T}T_y)=0$.
The inequality (\ref{inequality1}) then can be re-expressed in terms of the expectation value of the operators $O_{1}$ as
\begin{eqnarray}
-2\lambda_0(\sqrt{2}\lambda_2+\sqrt{2}\lambda_3+\lambda_1) &\leq& \langle O_1\rangle_{\psi_{\lambda_{1},\lambda_{2},\lambda_{3}}}\nonumber\\&-&
\frac{\langle\langle O_1\rangle_{\psi_{\lambda_{1},\lambda_{2},\lambda_{3}}}\rangle^{2}}{4}
\label{inequality1231}
\end{eqnarray}
If $\sqrt{2}\lambda_2+\sqrt{2}\lambda_3+\lambda_1=0$, then RHS of inquality (\ref{inequality1231}) is always positive irrespective of the values of the state parameter $(\lambda_1,\lambda_2,\lambda_3)$ . Thus it is not possible to differentiate between the class of states $|\psi_{\lambda_{1},\lambda_{2},\lambda_{3}}\rangle \in S_{4}$ and the class of states $|\psi_{\lambda_{1}=0,\lambda_{2}=0,\lambda_{3}=0}\rangle \in S_{1}$.\\

\textbf{Case-II:}
$\mu_{max}({T_x}^{T}T_x)=2{\lambda_0}^2(1-\lambda_0^{2}+\sqrt{-4(\lambda_2\lambda_3-\lambda_1\lambda_4)^2+(1-\lambda_0^{2})^{2}})$ and $\mu_{min}({T_y}^{T}T_y)=0$
Then the inequality (\ref{inequality1}) can be re-written as
\begin{eqnarray}
-2\lambda_0(\sqrt{2}\lambda_2+\sqrt{2}\lambda_3+\lambda_1) &\leq& \langle O_1\rangle_{\psi_{\lambda_{1},\lambda_{2},\lambda_{3}}}-P_4
\label{inequality1232}
\end{eqnarray}
where,
\begin{eqnarray}
P_4&=&2{\lambda_0}^2(1-\lambda_0^2\nonumber\\&+&\sqrt{-4(\lambda_2\lambda_3-\lambda_1\lambda_4)^2+(1-\lambda_0^{2})^{2}})
\nonumber\\&=& 2 \langle O_{5}\rangle_{\psi_{\lambda_{1},\lambda_{2},\lambda_{3}}}(1-\langle O_{5}\rangle_{\psi_{\lambda_{1},\lambda_{2},\lambda_{3}}}\nonumber\\&+&\sqrt{-\frac{\langle O_{6}\rangle_{\psi_{\lambda_{1},\lambda_{2},\lambda_{3}}}^{2}}{4}+(1-\langle O_{5}\rangle_{\psi_{\lambda_{1},\lambda_{2},\lambda_{3}}})^{2}})
\end{eqnarray}
We can now define an Hermitian operator $H_{7}$ as
\begin{eqnarray}
H_{7}&=&O_1-P_4I
\label{h7}
\end{eqnarray}
Therefore, the inequality (\ref{inequality1232}) can be re-formulated as
\begin{eqnarray}
-2\lambda_0(\sqrt{2}\lambda_2+\sqrt{2}\lambda_3+\lambda_1) \leq \langle H_{7}\rangle_{\psi_{\lambda_{1},\lambda_{2},\lambda_{3}}}
\label{inequality1233}
\end{eqnarray}
If $-2\lambda_0(\sqrt{2}\lambda_2+\sqrt{2}\lambda_3+\lambda_1)=0$ then $\langle H_{7}\rangle_{\psi_{\lambda_{1},\lambda_{2},\lambda_{3}}}\geq 0$ for all states $|\psi_{\lambda_{1}=0,\lambda_{2}=0,\lambda_{3}=0}\rangle \in S_{1}$.\\
For $-2\lambda_0(\sqrt{2}\lambda_2+\sqrt{2}\lambda_3+\lambda_1)\neq0$, we can calculate $Tr(H_{7}\rho_{\lambda_{1},\lambda_{2},\lambda_{3}})$ which is given by
\begin{eqnarray}
Tr(H_{7}\rho_{\lambda_{1},\lambda_{2},\lambda_{3}})&=&4\lambda_0\lambda_4-2{\lambda_0}^2({\lambda_1}^2+{\lambda_2}^2+{\lambda_3}^2+{\lambda_4}^2
\nonumber\\&+&\sqrt{T_4})
\label{tr13}
\end{eqnarray}
where,
\begin{eqnarray}
T_4&=&{\lambda_1}^4+{\lambda_2}^4+{\lambda_3}^4+{\lambda_4}^4+8\lambda_1\lambda_2\lambda_3\lambda_4-2{\lambda_2}^2{\lambda_3}^2+
\nonumber\\&&2{\lambda_2}^2{\lambda_4}^2+2{\lambda_1}^2{\lambda_2}^2+2{\lambda_1}^2{\lambda_3}^2-2{\lambda_1}^2{\lambda_4}^2
\nonumber\\&&+2{\lambda_3}^2{\lambda_4}^2
\end{eqnarray}
It can be easily shown that there exist state parameters $(\lambda_0,\lambda_1, \lambda_2, \lambda_3, \lambda_4)$ for which $Tr(H_{7}\rho_{\lambda_{1},\lambda_{2},\lambda_{3}})<0$. For instance, if we take $\lambda_0=0.6$, $\lambda_1=0.785812$, $\lambda_2=0.1$, $\lambda_3=0.05$ and $\lambda_4=0.1$, we get $Tr[H_7\rho_{\lambda_{123}}]=-0.303798$. Thus, the classification witness operator $H_7$ classify the
class of states described by the density operator $\rho_{\lambda_{1},\lambda_{2},\lambda_{3}} \in S_{4}$ and the class of states described by $\rho_{\lambda_{1}=0,\lambda_{2}=0,\lambda_{3}=0} \in S_{1}$

\section{Examples}
In this section, we have provided few examples of three-qubit states for which we construct classification witness operators.\\
\textbf{Example-1:} The three-qubit maximal slice state is given by \cite{ghoses},
\begin{eqnarray}
|MS\rangle=\frac{1}{\sqrt{2}}(|000\rangle+c|110\rangle+d|111\rangle),~~ c^2+d^2=1
\label{MS}
\end{eqnarray}
Let us consider the classification witness operators $H_{1}$, $H_{2}$ and $H_{3}$. The classification witness operator $H_{3}$ for (\ref{MS}) is now reduces to
\begin{eqnarray}
W_{MS}=O_1-I
\end{eqnarray}
The expectation value of $W_{MS}$ with respect to the state $|MS\rangle$ can be evaluated as
\begin{eqnarray}
Tr(W_{MS}\rho_{MS})=(2d-1)
\end{eqnarray}
Therefore, we can verify that $Tr(W_{MS}|MS\rangle\langle MS|)<0$ for the state parameter $d<\frac{1}{2}$. Further, it is easy to verify that the expectation value of the witness operator $H_{3}$ is positive for all state belong to subclass-I. Since the given state is detected by the classification witness operator $H_{3}$ so state (\ref{MS}) belongs to subclass-II. To investigate the form of the given state lying within subclass-II, we need to further classify it from the other classes of states belong to subclass-II. We can check that in the same range of the state parameter $d$ i.e. for $d<\frac{1}{2}$, the value of $Tr(H_1\rho_{MS})$ and $Tr(H_2\rho_{MS})$ are non-negative. Thus, we can say that the classification witness operators $H_3$ discriminate the maximal slice state from subclass-I and also it detects the state in the form (\ref{psi3}).\\
\textbf{Example-2:} Let us consider another three-qubit state defined as
\begin{eqnarray}
|\phi\rangle=\sqrt{p}|G\rangle-\sqrt{1-p}|K\rangle
\label{ex2state}
\end{eqnarray}
where,
\begin{eqnarray}
|G\rangle&=&a|000\rangle+b|111\rangle,~~a^{2}+b^{2}=1\nonumber\\
|K\rangle&=&c|110\rangle+d|101\rangle,~~c^{2}+d^{2}=1
\end{eqnarray}
 Now our task is to construct classification witness operator that may distinguish it from the state belong to subclass-I and also detect the form of the given state that belong to a particular class within subclass-III. To accomplish our task, let us consider classification witness operators $H_{4}$, $H_{5}$ and $H_{6}$ given in (\ref{h4}) and (\ref{h5}). We find that the expectation value of the witness operator $H_{4}$ is positive for all state belong to subclass-I but it gives negative value for some states belong to subclass-III. Hence the state (\ref{ex2state}) belongs to subclass-III. Moreover, we have investigated this classification problem within the subclass-III by constructing a table below. It shows that the expectation value of classification witness operator $H_{4}$ is negative for some range of the state parameter $p$ while the expectation value of other classification witness operators $H_{5}$ and $H_{6}$ gives positive values for the same range of the state parameters. This means that the given state (\ref{ex2state}) belong to subclass-III and it takes the form (\ref{psi23}).
In this table we have found the range of p where the witness operator $H_4$ detects the GHZ state given in the example whereas $H_5$ and $H_6$ do not detect the given GHZ state.
\begin{table}
	\begin{center}
		\caption{Range of the parameter p for which the classification witness operator $H_4$ classify the given GHZ state within the subclass-III}
		\begin{tabular}{|c|c|c|c|c|}\hline
			State parameter & p & $Tr[H_5{\rho}]$ & $Tr[H_4{\rho}]$ & $Tr[H_6{\rho}]$ \\ \hline (a, c) & & & \\  \hline
			(0.8, 0.3)  & (.291,.3) & $>0$ & $<0$ & $>0$ \\\hline
			(0.9, 0.4)  & (.548,.57) & $>0$ & $<0$ & $>0$ \\\hline
			(0.91, 0.8)  & (.4,.51) & $>0$ & $<0$ & $>0$ \\\hline
			(0.85, 0.35)  & (.43,.45) & $>0$ & $<0$ & $>0$ \\\hline
			(0.88, 0.8)  & (.25,.385) & $>0$ & $<0$ & $>0$ \\\hline
			(0.78, 0.3)  & (.208,.22) & $>0$ & $<0$ & $>0$ \\\hline
			(0.95, 0.4)  & (.69,.7) & $>0$ & $<0$ & $>0$ \\\hline
			(0.83, 0.45)  & (.26,.31) & $>0$ & $<0$ & $>0$ \\\hline
		\end{tabular}
	\end{center}
\end{table}



\section{Conclusion}
To summarize, we have defined systematically different subclasses of pure three-qubit GHZ class. The subclass-I denoted by $S_{1}$ contain the states of the form $\lambda_{0}|000\rangle + \lambda_{1}|111\rangle$. In particular, if $\lambda_{0}=\lambda_{1}=\frac{1}{\sqrt{2}}$ then the three-qubit state reduces to standard GHZ state and it is known that this state is very useful in various quantum information processing task. In this work, it has been shown that there exist states either belong to subclass-II denoted by $S_{2}$ or subclass-III denoted by $S_{3}$ or subclass-IV denoted by $S_{4}$, that may be more useful in some teleportation scheme in comparison to the states belong to $S_{1}$. This observation gives the motivation to discriminate the states belong to $S_{i},i=2,3,4$ from the family of states belong to $S_{1}$. We have prescribed the method for the construction of the witness operator to study the classification of the states belong to $S_{i},i=2,3,4$. Later, we have supported our work with few examples.

\section{Acknowledgement}
A.K. would like to acknowledge the financial support
from CSIR. This work is supported by CSIR File No.
08/133(0027)/2018-EMR-1.

\section{Appendix}
\subsection{Appendix-1}
The Hermitian matrices $T_{x}^{T}T_{x}$ and $T_{y}^{T}T_{y}$ for the state $\rho_{\lambda_2}=|\psi_{\lambda_2 }\rangle \langle \psi_{\lambda_2}|$ are given by
\begin{eqnarray}
T_{x}^{T}T_{x}= T_{y}^{T}T_{y}=
\begin{pmatrix}
4\lambda_0^{2}\lambda_4^{2} & 0 & 4\lambda_0^{2}{\lambda_2}{\lambda_4}\\
0 & 4\lambda_0^{2}\lambda_4^{2} & 0\\
4{\lambda_0}^2{\lambda_2}{\lambda_4} & 0 & 4\lambda_0^{2}\lambda_2^{2}\\
\end{pmatrix}
\end{eqnarray}
The Hermitian matrices $T_{x}^{T}T_{x}$ and $T_{y}^{T}T_{y}$ for the state $\rho_{\lambda_3}=|\psi_{\lambda_3 }\rangle \langle \psi_{\lambda_3}|$ is given by
\begin{eqnarray}
T_{x}^{T}T_{x}=
\begin{pmatrix}
4\lambda_0^{2}(\lambda_3^{2}+\lambda_4^{2}) & 0 & 0\\
0 & 4\lambda_0^{2}\lambda_4^{2} & 0\\
0 & 0 & 0\\
\end{pmatrix}
\end{eqnarray}
\begin{eqnarray}
T_{y}^{T}T_{y}=
\begin{pmatrix}
4\lambda_0^{2}\lambda_4^{2} & 0 & 0\\
0 & 4\lambda_0^{2}(\lambda_3^{2}+\lambda_4^{2}) & 0\\
0 & 0 & 0\\
\end{pmatrix}
\end{eqnarray}
The maximum eigenvalue of $T_{x}^{T}T_{x}$ for the state $|\psi_{\lambda_i}\rangle (i=2,3)$  is given by
\begin{eqnarray}
\mu_{max}({T_x}^{T}T_x)=4\lambda_0^{2}(\lambda_i^{2}+\lambda_4^{2}), i=2,3
\label{maxeigenval1}
\end{eqnarray}
The minimum eigenvalue of $T_{y}^{T}T_{y}$ for the state $|\psi_{\lambda_i}\rangle (i=2,3)$ is given by
\begin{eqnarray}
\mu_{min}({T_y}^{T}T_y)=0
\label{mineigenval1}
\end{eqnarray}
\subsection{Appendix-2}
The Hermitian matrices $T_{x}^{T}T_{x}$ and $T_{y}^{T}T_{y}$ for the state $\rho_{\lambda_{2},\lambda_{3}}=|\psi_{\lambda_{2},\lambda_{3}}\rangle \langle \psi_{\lambda_{2},\lambda_{3}}|$ is given by,
\begin{eqnarray}
T_{x}^{T}T_{x}=
\begin{pmatrix}
4\lambda_0^{2}(\lambda_3^{2}+\lambda_4^{2}) & 0 & 4\lambda_0^{2}\lambda_2\lambda_4\\
0 & 4\lambda_0^{2}\lambda_4^{2} & 0\\
4\lambda_0^{2}\lambda_2\lambda_4 & 0 & 4\lambda_0^{2}\lambda_2^{2}\\
\end{pmatrix}
\end{eqnarray}
\begin{eqnarray}
T_{y}^{T}T_{y}=
\begin{pmatrix}
4\lambda_0^{2}\lambda_4^{2} & 0 & 4\lambda_0^{2}\lambda_2\lambda_4\\
0 & 4\lambda_0^{2}(\lambda_3^{2}+\lambda_4^{2}) & 0\\
4\lambda_0^{2}\lambda_2\lambda_4 & 0 & 4\lambda_0^{2}\lambda_2^{2}\\
\end{pmatrix}
\end{eqnarray}
The maximum eigenvalue of $T_{x}^{T}T_{x}$ is given by
\begin{eqnarray}
\mu_{max}({T_x}^{T}T_x)&=&max\{u,v_{1}\}
\label{maxeigenval23}
\end{eqnarray}
where $u=4\lambda_0^{2}\lambda_4^{2}$ and $v_{1}=2\lambda_0^{2}(1-\lambda_0^{2}+\sqrt{-4\lambda_2^2\lambda_3^{2}+(1-\lambda_0^{2})^2})$.\\
The minimum eigenvalue of $T_{y}^{T}T_{y}$ is given by
\begin{eqnarray}
\mu_{min}({T_y}^{T}T_y)=0
\label{mineigenval1}
\end{eqnarray}
\subsection{Appendix-3}
The Hermitian matrices $T_{x}^{T}T_{x}$ and $T_{y}^{T}T_{y}$ for the state $\rho_{\lambda_{1},\lambda_{2}}=|\psi_{\lambda_{1},\lambda_{2}}\rangle \langle \psi_{\lambda_{1},\lambda_{2}}|$ is given by

\begin{eqnarray}
T_{x}^{T}T_{x}=
\begin{pmatrix}
4\lambda_0^{2}\lambda_4^{2} & 0 & 4\lambda_0^{2}\lambda_2\lambda_4\\
0 & 4\lambda_0^{2}\lambda_4^{2} & 0\\
4\lambda_0^{2}\lambda_2\lambda_4 & 0 & 4\lambda_0^{2}(\lambda_1^{2}+\lambda_2^{2})\\
\end{pmatrix}
\end{eqnarray}
\begin{eqnarray}
T_{y}^{T}T_{y}=
\begin{pmatrix}
4\lambda_0^{2}\lambda_4^{2} & 0 & 4\lambda_0^{2}\lambda_2\lambda_4\\
0 & 4\lambda_0^{2}\lambda_4^{2} & 0\\
4\lambda_0^{2}\lambda_2\lambda_4 & 0 & 4\lambda_0^{2}\lambda_2^{2}\\
\end{pmatrix}
\end{eqnarray}
The Hermitian matrices $T_{x}^{T}T_{x}$ and $T_{y}^{T}T_{y}$ for the state $\rho_{\lambda_{1,3}}=|\psi_{\lambda_{1,3}}\rangle \langle \psi_{\lambda_{1,3}}|$ is given by
\begin{eqnarray}
T_{x}^{T}T_{x}=
\begin{pmatrix}
4\lambda_0^{2}(\lambda_3^{2}+\lambda_4^{2}) & 0 & 4\lambda_0^{2}\lambda_1\lambda_3\\
0 & 4\lambda_0^{2}\lambda_4^{2} & 0\\
4\lambda_0^{2}\lambda_1\lambda_3 & 0 & 4\lambda_0^{2}\lambda_1^{2}\\
\end{pmatrix}
\end{eqnarray}
\begin{eqnarray}
T_{y}^{T}T_{y}=
\begin{pmatrix}
4\lambda_0^{2}\lambda_4^{2} & 0 & 0\\
0 & 4\lambda_0^{2}(\lambda_3^{2}+\lambda_4^{2}) & 0\\
0 & 0 & 0\\
\end{pmatrix}
\end{eqnarray}
The maximum eigenvalue of $T_{x}^{T}T_{x}$ is given by
\begin{eqnarray}
\mu_{max}({T_x}^{T}T_x)&=&max\{u,v_{i}\}, i=2,3
\label{maxeigenval23}
\end{eqnarray}
where $v_{i}=2\lambda_0^{2}(\lambda_1^{2}+\lambda_{i}^{2}+\lambda_4^{2}+\sqrt{(\lambda_1^{2}+\lambda_{i}^{2}+\lambda_4^{2})^2-4\lambda_1^2\lambda_4^{2}})$,~i=2,3.\\
The minimum eigenvalue of $T_{y}^{T}T_{y}$ is given by
\begin{eqnarray}
\mu_{min}({T_y}^{T}T_y)=0
\label{mineigenval1}
\end{eqnarray}
\subsection{Appendix-4}
The Hermitian matrices $T_{x}^{T}T_{x}$ and $T_{y}^{T}T_{y}$ for the state $\rho_{\lambda_{1},\lambda_{2},\lambda_{3}}=|\psi_{\lambda_{1},\lambda_{2},\lambda_{3}}\rangle \langle \psi_{\lambda_{1},\lambda_{2},\lambda_{3}}|$ is given by

\begin{eqnarray}
T_{x}^{T}T_{x}=
\begin{pmatrix}
4\lambda_0^{2}(\lambda_3^{2}+\lambda_4^{2}) & 0 & 4\lambda_0^{2}(\lambda_1\lambda_3+\lambda_2\lambda_4)\\
0 & 4\lambda_0^{2}\lambda_4^{2} & 0\\
4\lambda_0^{2}(\lambda_1\lambda_3+\lambda_2\lambda_4) & 0 & 4\lambda_0^{2}(\lambda_1^{2}+\lambda_2^{2})\\
\end{pmatrix}
\end{eqnarray}
\begin{eqnarray}
T_{y}^{T}T_{y}=
\begin{pmatrix}
4\lambda_0^{2}\lambda_4^{2} & 0 & 4\lambda_0^{2}\lambda_2\lambda_4\\
0 & 4\lambda_0^{2}(\lambda_3^{2}+\lambda_4^{2}) & 0\\
4\lambda_0^{2}\lambda_2\lambda_4 & 0 & 4\lambda_0^{2}\lambda_2^{2}\\
\end{pmatrix}
\end{eqnarray}
The maximum eigenvalue of $T_{x}^{T}T_{x}$ is given by
\begin{eqnarray}
\mu_{max}({T_x}^{T}T_x)&=&max\{u,v_{4}\}
\label{maxeigenval23}
\end{eqnarray}
where $v_{4}=2{\lambda_0}^2(k+\sqrt{-4(\lambda_2\lambda_3-\lambda_1\lambda_4)^2+k^{2}})$,\\ $k=1-\lambda_0^{2}$.\\
The minimum eigenvalue of $T_{y}^{T}T_{y}$ is given by
\begin{eqnarray}
\mu_{min}({T_y}^{T}T_y)=0
\label{mineigenval123}
\end{eqnarray}
\subsection{Appendix-5}
\textbf{Classification witness operator for the classification of states contained in subclass-II and subclass-III}\\
\\
The GHZ class of state within subclass-III with state parameters ($\lambda_0$, $\lambda_1$, $\lambda_2$, $\lambda_4$) is given by
\begin{eqnarray}
|\psi_{\lambda_{1},\lambda_{2}}\rangle=\lambda_0|000\rangle+\lambda_1|100\rangle+\lambda_2|101\rangle+\lambda_4|111\rangle
\label{psi12}
\end{eqnarray}
with $\lambda_0^{2}+\lambda_1^{2}+\lambda_2^{2}+\lambda_4^{2}=1$.\\
The Hermitian matrices $T_{x}^{T}T_{x}$ and $T_{y}^{T}T_{y}$ for the state $\rho_{\lambda_{1},\lambda_{2}}=|\psi_{\lambda_{1},\lambda_{2}}\rangle \langle \psi_{\lambda_{1},\lambda_{2}}|$ are given in appendix-3.\\
The expression for $Tr[({T_x}^{T}T_x)+({T_y}^{T}T_y)]$ is given by
\begin{eqnarray}
Tr[({T_x}^{T}T_x)+({T_y}^{T}T_y)]&=&16\lambda_0^{2}\lambda_4^{2}+8\lambda_0^{2}\lambda_2^{2}+\nonumber\\&&
4\lambda_0^{2}\lambda_1^{2}
\label{Tracecond131}
\end{eqnarray}

\textbf{Case-I:} If $\mu_{max}({T_x}^{T}T_x)=4\lambda_0^{2}\lambda_4^{2}$ and $\mu_{min}({T_y}^{T}T_y)=0$.
The inequality (\ref{inequality1}) then can be re-expressed in terms of the expectation value of the operators $O_{1}$ as
\begin{eqnarray}
-2\lambda_0(\sqrt{2}\lambda_2+\lambda_1) &\leq& \langle O_1\rangle_{\psi_{\lambda_1,\lambda_2}}-\frac{(\langle O_1\rangle_{\psi_{\lambda_1,\lambda_2}})^{2}}{4}
\end{eqnarray}
If $\lambda_2=0$, then above inequality becomes,
\begin{eqnarray}
-2\lambda_0\lambda_1 \leq \langle O_1\rangle_{\psi_{\lambda_1,\lambda_2}}-\frac{(\langle O_1\rangle_{\psi_{\lambda_1,\lambda_2}})^{2}}{4}
\label{inequality121}
\end{eqnarray}
The R.H.S of the inequality is positive for every state $|\psi\rangle$. Thus it is not possible to differentiate between the class of states $|\psi_{\lambda_{1},\lambda_{2}=0}\rangle \in S_{2}$ and $|\psi_{\lambda_{1},\lambda_{2}}\rangle \in S_{3}$, using the inequality (\ref{inequality121}) for this case.\\
\textbf{Case-II:}
If $\mu_{max}({T_x}^{T}T_x)=2\lambda_0^{2}(\lambda_1^{2}+\lambda_{i}^{2}+\lambda_4^{2}+\sqrt{(\lambda_1^{2}+\lambda_{i}^{2}+\lambda_4^{2})^2-4\lambda_1^2\lambda_4^{2}})$ and $\mu_{min}({T_y}^{T}T_y)=0$
Then the inequality (\ref{inequality1}) can be re-written as
\begin{eqnarray}
-2\lambda_0(\sqrt{2}\lambda_{2}+\lambda_1) &\leq& \langle O_1\rangle_{\psi_{\lambda_1,\lambda_i}}-P_{2}
\label{inequality122}
\end{eqnarray}
We can now define an Hermitian operator $H_{8}$, as
\begin{eqnarray}
H_{8}&=& O_1-P_2I+\frac{O_4}{2}
\label{h7}
\end{eqnarray}
Therefore, the inequality (\ref{inequality122}) can be re-formulated as
\begin{eqnarray}
-2\sqrt{2}\lambda_0\lambda_2
\leq \langle H_{8}\rangle_{\psi_{\lambda_1,\lambda_2}}
\label{inequality123}
\end{eqnarray}
If $\lambda_2=0$ then $\langle H_{8}\rangle_{\psi_{\lambda_1,\lambda_2}} \geq 0$, for all states $|\psi_{\lambda_{1}=0,\lambda_{2}=0} \rangle \in S_{1}$ and $|\psi_{\lambda_{1},\lambda_{2}=0}\rangle \in S_{2}$.\\
For $\lambda_2 \neq 0$, then there exist state parameters $(\lambda_0,\lambda_1, \lambda_{2}, \lambda_4)$ for which $Tr(H_{8}\rho_{\lambda_1,\lambda_{2}})<0$. For instance, if we take $\lambda_0=0.01$, $\lambda_1=0.948631$, $\lambda_2=0.3$ and $\lambda_4=0.1$, we get $Tr[H_8\rho_{\lambda_{1},\lambda_{2}}]=-0.129027$. Thus, the Hermitian operator $H_{8}$ serves as a classification witness operator and classify GHZ class of states described by the density operator $\rho_{\lambda_{1},\lambda_{2}} \in S_{3}$ and the GHZ of states $\rho_{\lambda_{1}=0,\lambda_{2}=0} \in S_{1}$ or $\rho_{\lambda_{1},\lambda_{i}=0},(i=2,3) \in S_{2}$.\\
Simillarly, we can construct witness operator that can classify GHZ states belonging to subclass-III and subclass-IV.


\begin{thebibliography}{90}
\bibitem{piani} R. Horodecki, P. Horodecki, M. Horodecki, and K. Horodecki,
Rev. Mod. Phys. \textbf{81}, \textbf{865} (2009);M. Piani, S. Gharibian, G. Adesso, J. Calsamiglia, P.
Horodecki,and A. Winter, Phys. Rev. Lett. \textbf{106}, 220403 (2011)
\bibitem{dur} W. Dur, G. Vidal, and J. I. Cirac, Phys. Rev. A \textbf{62}, 062314 (2000).
\bibitem{datta} C. Datta, S. Adhikari, A. Das, and P. Agrawal, Eur. Phys. J. D \textbf{72}, 157 (2018).
\bibitem{singh} A. Singh, H. Singh, K. Dorai, and Arvind, Phys. Rev. A \textbf{98}, 032301 (2018).
\bibitem{acin} A. Acin, D. Bruss, M. Lewenstein and A. Sanpera, Phys. Rev. Lett. \textbf{87}, 040401 (2001).
\bibitem{sabin} C. Sabin and G. Garcia-Alcaine, Eur. Phys. J. D \textbf{48}, 435 (2008).
\bibitem{bera} M. N. Bera, R. Prabhu, A. Sen(De), and U. Sen, Phys. Rev. A \textbf{86}, 012319 (2012).
\bibitem{verstraete} F. Verstraete, J. Dehaene, B. De Moor, and H. Verschelde, Phys. Rev. A \textbf{65}, 052112 (2002).
\bibitem{viehmann} O. Viehmann, C. Eltschka, and J. Siewert, Phys. Rev. A \textbf{83}, 052330 (2011).
\bibitem{zangi} S. M. Zangi, J-L Li, and C-F Qiao, J. Phys. A: Math. Theor. \textbf{50}, 325301 (2017).
\bibitem{miyake} A. Miyake, Phys. Rev. A \textbf{67}, 012108 (2003).
\bibitem{chen} L. Chen and Y-X Chen, Phys. Rev. A \textbf{74}, 062310 (2006).
\bibitem{li1} X. Li and D. Li, Phys. Rev. Lett. \textbf{108}, 180502 (2012).
\bibitem{miyake1} A. Miyake and M. Wadati, Quant. Info. Comp. \textbf{2} (Special), 540 (2002).
\bibitem{acin1} A. Acin, A. Andrianov, L. Costa, E. Jane, J. I. Lattore, and R. Tarrach, Phys. Rev. Lett. \textbf{85}, 1560 (2000).
\bibitem{coffman}V. Coffman, J. Kundu, and W. K. Wootters, Phys. Rev. A 61, 052306 (2000).
\bibitem{soojoonlee}S. Lee,J. Joo and J. Kim, Phys. Rev. A 72, 024302 (2005).
\bibitem{ghoses} S. Ghose, N. Sinclair, S. Debnath, P. Rungta and R. Stock, Phys. Rev. Lett. \textbf{102}, 250404 (2009).
\bibitem{adhikari} S. Adhikari, Journal of Experimental and Theoretical Physics 131, 375 (2020).
\bibitem{sudbery} A. Sudbery, J. Phys. A: Math. Gen. 34 643 (2001).
\bibitem{horn}R. A. Horn, and C. R. Johnson, Matrix analysis, (Cambridge University Press, Cambridge, 1999).







\end{thebibliography}
\end{document}